\documentclass[ aip, jmp, amsmath,amssymb, reprint, ]{revtex4-1}
\usepackage{graphicx}
\usepackage{dcolumn}
\usepackage{bm}
\usepackage{amssymb}
\usepackage{amsmath}
\usepackage{amsfonts}

\usepackage{hyperref}
\usepackage{bbm}
\usepackage[usenames]{color}
\begin{document}

\preprint{AIP/123-QED}

\title{Andronov--Hopf bifurcation with and without parameter in a cubic memristor oscillator with a line of equilibria}

\author{Ivan A. Korneev}
\affiliation{Department of Physics, Saratov State University, Astakhanskaya str. 83, 410012 Saratov, Russia}
\author{Vladimir V. Semenov}
\email{semenov.v.v.ssu@gmail.com}
\affiliation{Department of Physics, Saratov State University, Astakhanskaya str. 83, 410012 Saratov, Russia}

\date{\today}

\begin{abstract}
The model of a memristor-based oscillator with cubic nonlinearity is studied. The considered system has infinitely many equilibrium points, which build a line of equilibria in the phase space. Numerical modeling of the dynamics is combined with bifurcational analysis. It is shown that oscillation excitation has distinctive features of the supercritical Andronov--Hopf bifurcation and can be achieved by changing of a parameter value as well as by variation of initial conditions. Therefore the considered bifurcation is called Andronov-Hopf bifurcation with and without parameter.
\end{abstract}

\maketitle

\begin{quotation}
Bifurcations without parameters are a frequent occurrence in  systems with a line of equilibria. Such bifurcations are induced by changing of the initial conditions, whereas parameter values are kept constant. In the present paper a bifurcation, which is determined both by initial conditions and a parameter value, is studied on the example of a memristor-based oscillator with cubic nonlinearity. Mechanism of soft oscillation excitation in the system under study is rigorously revealed. It is determined by the memristor characteristic and has distinctive features of the supercritical Andronov--Hopf bifurcation. It has been shown that the intrinsic peculiarity of the system is the continuous functional dependence of the oscillation amplitude both on the parameter value and on the initial conditions. 
\end{quotation}

\section{Introduction}
The two-terminal passive electronic circuit element "memristor" was introduced by Leon Chua \cite{chua1971} as realization of relationship between the electrical charge and the magnetic flux linkage. Then the term "memristor" was extended to the conception of "memristive systems" \cite{chua1976}.  A class of the memristive systems is identified by a continuous functional dependence of characteristics at any time on previous states. The memristive systems attract attention due to their potential applications in electronics and neuroscience \cite{kozma2012,adamatzky2014,tetzlaff2014,radwan2015,vourkas2016,vaidyanathan2017,ventra2013,li2013}. 

Memristor intrinsic properties can essentially change the dynamics of electronic oscillatory systems and induce qualitatively new types of the behavior.  For this reason the memristor is interesting for specialists in nonlinear dynamics.
There are well-known memristor-based chaotic \cite{itoh2008,buscarino2012,buscarino2013,pham2013,gambuzza2015-2} and Hamiltonian \cite{itoh2011,itoh2017} oscillators. The existence of hidden attractors in systems including memristor has been revealed \cite{pham2015,chen2015,chen2015-2}. The single memristor under the summary influence of noise and periodic signal exhibits a stochastic resonance phenomenon \cite{stotland2012}. Memristor topics address the collective dynamics of coupled through the memristor oscillators and ensembles of coupled memristor-based oscillators: from synchronization of two chaotic or regular self-sustained oscillators coupled through the memristor \cite{frasca2014,volos2015,volos2016,frasca2015,ignatov2016} to
spatio-temporal phenomena in ensembles of interacting oscillators with adaptive coupling realized by the memristor or in ensembles of coupled memristor-based oscillators \cite{pham2012,gambuzza2015,buscarino2016,zhao2015}. 

Memristor-based oscillators with manifolds of equilibria represent a special class of dynamical systems \cite{itoh2008,messias2010,botta2011,riaza2012}. The m-dimensional manifolds of equilibria consist of non-isolated equilibrium points. In the simplest case this manifold can exist such as a line of equilibria (m = 1). Normally hyperbolic manifolds of equilibria are distinguished and their points are characterized by m purely imagine or zero eigenvalues, whereas all the other eigenvalues have nonzero real parts. In terms of the dynamical system theory, the systems whose phase space includes the normally hyperbolic manifolds of equilibria can be referred to a special kind of systems with unusual characteristics. This class of dynamical systems has been considered mathematically  \cite{fiedler2000-1,fiedler2000-2,fiedler2000-3,liebscher2015,riaza2012,riaza2016}. It has been shown that their significant property is the existence of so-called bifurcations without parameters, i.e., the bifurcations corresponding to fixed parameters when the condition of normal hyperbolicity is violated at some points of the manifold of equilibria. 

Bifurcational mechanisms of oscillation excitation in the proposed by L.Chua and M.Itoh \cite{itoh2008} memristor-based system with a line of equilibria were described in details \cite{korneev2017}. Equations of the dynamical model including the memristor with a piecewise-smooth characteristic were analytically solved using the quasiharmonical reduction. Depending on nonlinearity of the system, hard or soft oscillation excitation can be realized. It has been shown that hard and soft excitation scenarios have principally different nature. In the system including only memristor piecewise-smooth nonlinearity, the hard oscillation excitation is a result of a border-collision bifurcation. In the system with additional nonlinear element the soft excitation is caused by impact of a smooth nonlinear function and has distinctive features of the supercritical Andronov--Hopf bifurcation. Specific character of the hard excitation in the considered system is caused by the piecewise-smooth Chua's memristor characteristic. This result raises the next question of how bifurcational mechanisms change in the presence of smooth nonlinearity of the memristor. One of the simplest models of the memristor with smooth characteristic is described by a cubic nonlinear function \cite{messias2010,chua2011,liu2015}. The dynamics of a system with a line of equilibria including the cubic memristor was studied by means of numerical simulation and linear stability analysis of equilibrium points \cite{messias2010}. Nevertheless, analysis of fixed point stability does not provide to uncover bifurcational mechanisms of oscillation excitation. It is necessary to solve model equations for complete analysis. In the present paper the model equations are analytically studied and full bifurcational analysis is carried out. As will be shown below, oscillation excitation mechanism is associated with the supercritical Andronov--Hopf scenario. Analytically derived formula for the periodic solution amplitude constitutes dependence both on a parameter value and on a initial condition. Therefore the considered type of bifurcation is called bifurcation with and without parameters.

\section{Model and Methods}
\label{mod_meth}

\begin{figure*}
\begin{center}
\includegraphics[width=0.9\columnwidth]{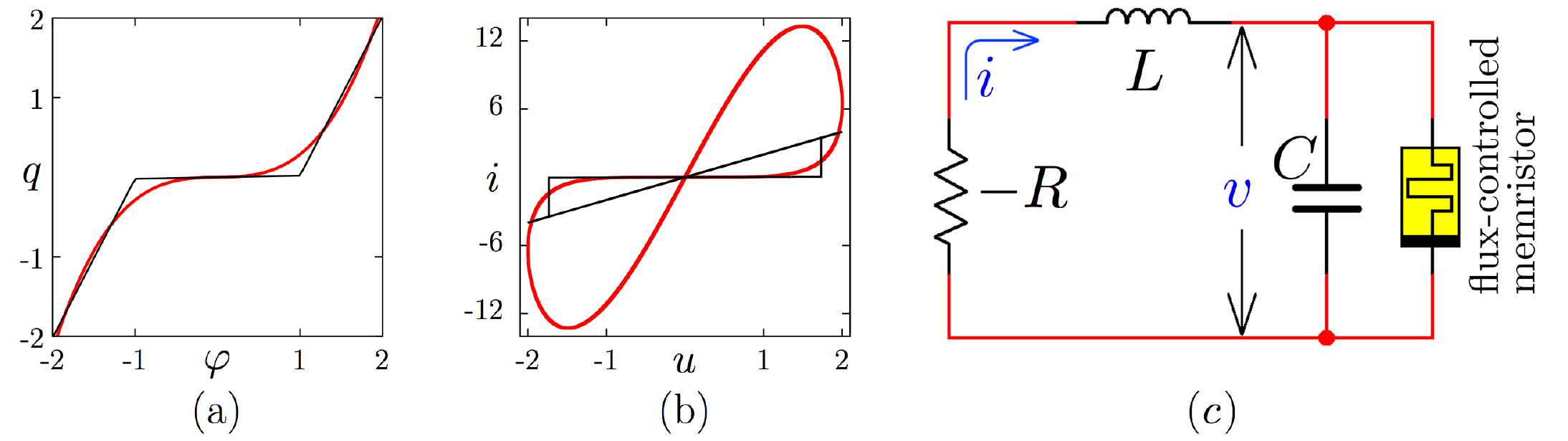}
\end{center}
\caption{(a) Dependence $q(\varphi)$ of piecewise-smooth Chua's memristor (\ref{q_phi_chua_memristor}) with parameters $a=0.02$, $b=2$, $\varphi_{0}=1$ (the black curve) and its approximation by function $q(\varphi)=0.02\varphi+\frac{0.8}{3}\varphi^{3}$ (the red curve). (b) Current-voltage characteristic of the memristor driven by periodic voltage signal $U_{ext}=2 \sin{(t)}$ corresponding to Chua's memristor (\ref{chua_memristor}) with parameters $a=0.02$, $b=2$, $\varphi_{0}=1$ (the black loop) and cubic memristor (\ref{cubic_memristor}) with parameters $a=0.02$, $b=0.8$ (the red loop). (c) Schematic circuit diagram of the system under study (Eqs.(\ref{system})).}
\label{figure1}                                                                                                   
\end{figure*}

The flux controlled memristor relates the transferred electrical charge, $q(t)$, and the magnetic flux linkage, $\varphi(t)$: $dq=Wd\varphi$, whence it follows that $W=W(\varphi)=\dfrac{dq}{d\varphi}$. By using the formulas $d\varphi=Udt$ and $dq=idt$ ($U$ is the voltage across the memristor, $i$ is the current passing through the memristor) the memristor current-voltage characteristic can be derived: $i=W(\varphi)U$. It means that $W$ is the flux-controlled conductance (memductance) and depends on the entire past history of $U(t)$:
 \begin{equation}
W(\varphi)=\dfrac{dq}{d\varphi}=q '\left( \int\limits_{-\infty}^{t}{U(t)dt} \right).
\label{W(phi)}
\end{equation}
Initially offered by Leon Chua flux-controlled memristor \cite{chua1971} is described by piecewise-linear dependence $q(\varphi)$ (Fig. \ref{figure1}(a)):
\begin{equation}
q(\varphi)=
\begin{cases}
	  (a-b)\varphi_{0}+b\varphi, & \varphi \ge \varphi_{0},\\
          a \varphi , & |\varphi| < \varphi_{0},\\
          -(a-b)\varphi_{0}+b\varphi , & \varphi \le -\varphi_{0}.
          
\end{cases}
\label{q_phi_chua_memristor}
\end{equation}

Then the memristor conductance $W(\varphi)$ becomes:

\begin{equation}
W(\varphi)=
\begin{cases}
          a , & |\varphi| < \varphi_{0},\\
          b , & |\varphi| \geq \varphi_{0}.
\end{cases}
\label{chua_memristor}
\end{equation}
This dependence $q(\varphi)$ (Eq. (\ref{q_phi_chua_memristor})) can be approximated by the cubic function $q(\varphi)=a\varphi+\frac{1}{3}b\varphi^{3}$ (Fig. \ref{figure1}(a)). Then the cubic memristor conductance $W(\varphi)$ is:
\begin{equation}
W(\varphi)=a+ b\varphi^{2}.
\label{cubic_memristor}
\end{equation}
Substitution of the cubic memristor $q-\varphi$  curve does not involve qualitative changes in memristor functioning. The classical loop in the current-voltage characteristic of the memristor driven by the external periodic influence, $U_{ext}=U_{0}sin(\omega_{ext}t)$, persists (Fig. \ref{figure1}(b)). The model (\ref{cubic_memristor}) will be used in the following.

The system under study is depicted in Fig. \ref{figure1}(c). It is the series oscillatory circuit including the element with constant negative resistance and the flux-controlled memristor. The presented in Fig. \ref{figure1}(c) system is described by the following dynamical variables: $v$ is the voltage across the capacitance $C$,  $i$ is the current through the inductance $L$, and the magnetic flux, $\varphi$, controlling the memristor. In the dimensionless form the model equations can be derived: 
\begin{equation}
\frac{dx}{dt}=\alpha (y-W(z)x), \frac{dy}{dt} = -\gamma x +\beta y , \frac{dz}{dt}=x,
\label{system}
\end{equation}
where $x \sim v$, $y \sim i$, $z \sim \varphi$ are the dimensionless variables, $\alpha \sim 1/C$ and $\gamma\sim 1/L$ are the dimensionless parameters, $W(z)=a+bz^{2}$ is the presented in the dimensionless form flux-controlled conductance of the cubic memristor. The parameter $\beta \sim R/L$ characterizes the element with negative resistance. The parameters $\alpha$ and $\gamma$ are set to be equal to unity. 

The system (\ref{system}) is explored both theoretically by using quasiharmonic reduction and numerically by means of modelling methods. Numerical simulations are carried out by integration of Eqs. (\ref{system}) using the Heun method \cite{mannella2002} with time step  $t = 0.0001$ from different initial conditions.

\section{Numerical simulations and theoretical analysis}
\label{num_mod}

Let us consider the system (\ref{system}). It is evident that the system (\ref{system}) has a line of equilibria, i.e., each point on the axis OZ is an equilibrium point. One of the eigenvalues $\lambda_{i}$ of the equilibria is always equal to zero and the others depend on the parameters and the position of the point on the OZ-axis (z coordinate):
\begin{equation}
\lambda_{1}=0, \\
\lambda_{2,3}=\frac{\beta-W(z)}{2} \pm \sqrt{\frac{(W(z)+\beta)^2}{4}-1}. 
\label{eigenvalues}
\end{equation}
Each point of the line of equilibria is neutrally stable in the OZ-axis direction. Hereinafter, using the terms "stable" or "unstable" point at the line of equilibria, we mean the behavior of trajectories (attraction or repelling) in the neighborhood of the equilibrium point, which is determined by the eigenvalues $\lambda_{2,3}$.

\begin{figure}
\begin{center}
\includegraphics[width=1.0\columnwidth]{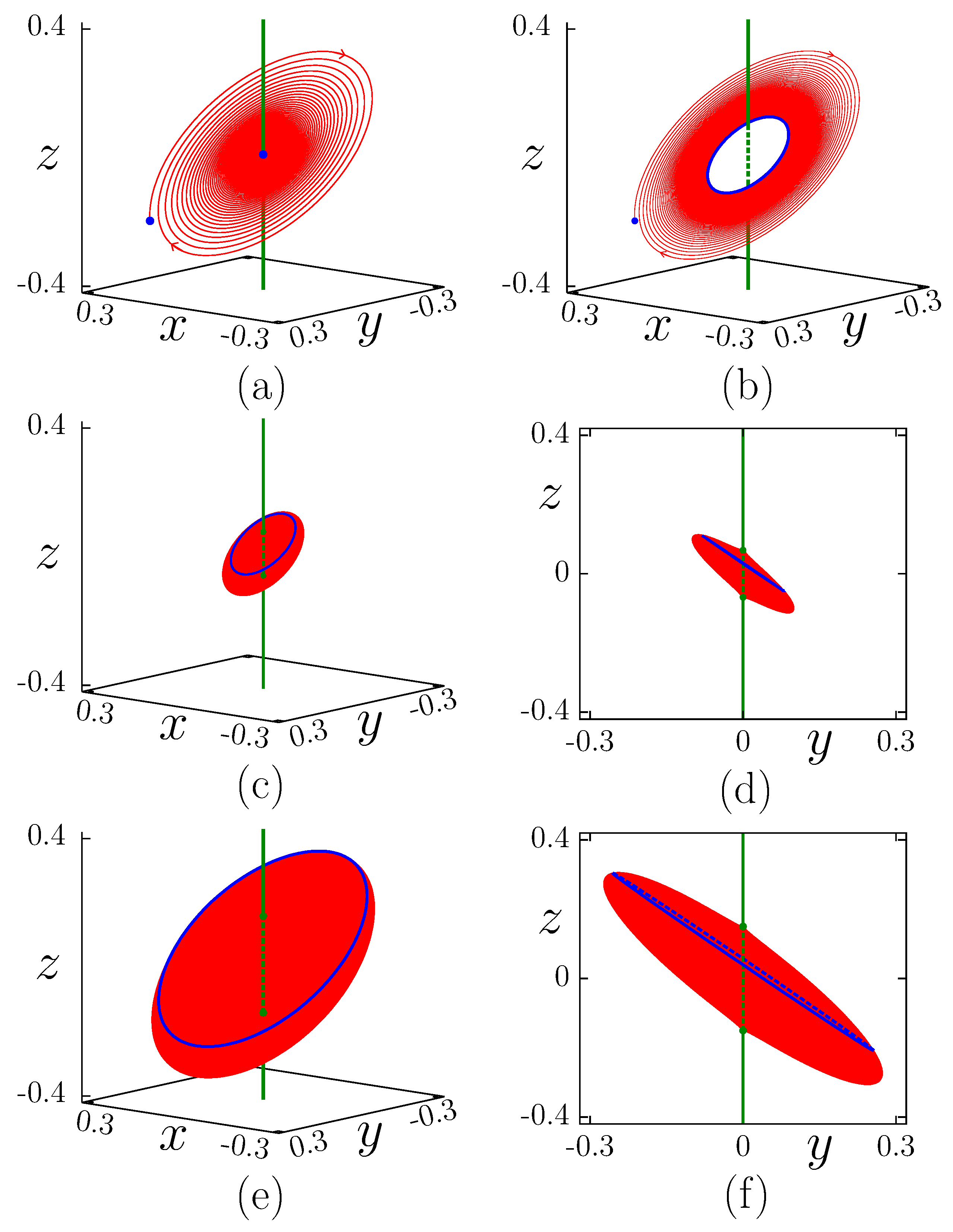}
\end{center}
\caption{Trajectories and attractor in the phase space of the system (\ref{system}): (a) motion to line of equlibria  (the trajectory is colored in red) for $\beta=0.01$; (b) motion to non-isolated closed curve (is colored in blue) for $\beta=0.022$; 
(c) Attractor corresponding to $\beta=0.022$ and its projection in plane ($y,z$) (the panel (d)); (e) Attractor corresponding to $\beta=0.035$ and its projection in plane ($y,z$) (the panel (f)). On all panels: a manifold of stable equilibria is shown by the green solid line, a manifold of unstable equilibria is shown by the green dashed line, points corresponding to initial conditions are shown by blue filled circles. On the panels (c)-(f): A two-dimensional surface formed by a manifold of invariant closed curves is colored in red. Invariant closed curve on the attractor is colored in blue. Other parameters are: $a=0.02$, $b=0.8$.}
\label{figure2}                                                                                                   
\end{figure}

Increasing of the parameter $\beta$ from zero gives rise to the following bifurcational changes in the phase space. In case $0\le \beta<a$, an attractor of the system is a manifold of stable equilibria, and all trajectories are attracted to him (Fig. \ref{figure2}(a)). When $\beta \ge a$, equilibrium points with coordinate $z\in\left(-\sqrt{\frac{\beta-a}{b}};\sqrt{\frac{\beta-a}{b}} \right)$ become unstable, while other points of the line of equilibria with  $|z|>\sqrt{\frac{\beta-a}{b}}$ remain to be stable. Starting from the vicinity of unstable point $(0;0;|z|<\sqrt{\frac{\beta-a}{b}})$ the phase point moves away from the initial state and traces a spiral-like trajectory. That movement culminates in motion along an invariant closed curve (the blue line in Fig. \ref{figure2}(b)). Any changes of the initial conditions give rise to hit on another invariant closed curve. In that way, structure of an attractor changes in the point $\beta=a$. After the bifurcation the attractor consists of a continuous set of closed curves which form a two-dimensional surface for $-\sqrt{\frac{\beta-a}{b}}<z<\sqrt{\frac{\beta-a}{b}}$ and of a set of points on the OZ-axis, for which $|z| \ge \sqrt{\frac{\beta-a}{b}}$ (Fig. \ref{figure2}(c)). The two-dimensional surface expands when the parameter $\beta$ increases (Fig. \ref{figure2}(c)-(f)). 

In order to understand reasons of attractor evolution, we transform the system (\ref{system}) to the oscillatory form:
\begin{equation}
\label{osc}
\left\lbrace
\begin{array}{l}
\dfrac{d^{2}x}{dt^{2}}-(\beta-a)\dfrac{dx}{dt}+(1-\beta a)x\\
\\
=b\beta z^{2}x-2bx^{2}z-bz^{2}\dfrac{dx}{dt}, \\
\\
\dfrac{dz}{dt}=x. \\
\end{array}
\right.
\end{equation}
Periodic solution of the first equation of the system (\ref{osc}) on the frequency $\omega_{0}=\sqrt{1-\beta a}$ is sought. In complex representation the solution $x(t)$ becomes:
\begin{equation}
\label{x}
\left.
\begin{array}{l}
x(t)=Re \left\{ A(t)\exp{(i\omega_{0}t)}\right\}=\dfrac{1}{2}( A \exp{(i \omega_{0}t)}\\
+A^{*} \exp{(-i \omega_{0}t)}) ,\\
\end{array}
\right.
\end{equation}
where $A(t)$ is the complex amplitude, $A^{*}(t)$ is the complex conjugate function. The following condition for the first derivative is assumed to be satisfied: $\dfrac{dA}{dt} \exp{(i\omega_{0}t)}+\dfrac{dA^{*}}{dt} \exp{(-i\omega_{0}t)}=0$. With taking this condition into consideration the first derivative becomes:
\begin{equation}
\label{dx}
\dfrac{dx}{dt}=\dfrac{1}{2}\left( i\omega_{0} A \exp {(i\omega_{0}t)} - i\omega_{0}A^{*} \exp {(-i\omega_{0}t)}    \right) .
\end{equation}
Then the second derivative is:
\begin{equation}
\label{ddx}
\left.
\begin{array}{l}
\dfrac{d^{2}x}{dt^{2}}= \left(i\omega_{0} \dfrac{dA}{dt}-\dfrac{\omega_{0}^{2}}{2}A\right) \exp{(i\omega_{0}t)} 
\\
-\left(i\omega_{0} \dfrac{dA^{*}}{dt}+\dfrac{\omega_{0}^{2}}{2}A^{*}\right) \exp{(-i\omega_{0}t)}.\\
\end{array}
\right.
\end{equation}
Using Eq. (\ref{x}) the variable $z(t)$ can be found as being:
\begin{equation}
\label{z}
\left.
\begin{array}{l}
z(t)\approx\dfrac{1}{2i\omega_{0}}( A \exp{(i \omega_{0}t)}\\ -A^{*} \exp{(-i \omega_{0}t)}) +z_{0},
\end{array}
\right.
\end{equation}
where $z_{0}$ is a constant determined by initial conditions. 

Substituting Eqs. (\ref{x})-(\ref{z}) into the first equation of the system (\ref{osc}), we approximate all fast oscillating terms by their averages over one period $T=2\pi/\omega_{0}$ which gives zero. Then we obtain the equation for the complex amplitude, $A$:

\begin{equation}
\label{A}
\left.
\begin{array}{l}
\dfrac{dA}{dt}=\dfrac{A}{2}\left( \beta-a-\dfrac{b|A|^{2}}{4\omega_{0}^{2}} -z_{0}^{2}b \right)\\
\\
-\dfrac{iAb\beta}{2\omega_{0}}\left( \dfrac{|A|^{2}}{4\omega_{0}^{2}} +z_{0}^{2}\right).
\end{array}
\right.
\end{equation}
Equation (\ref{A}) can be rewritten as a system of equations for the real amplitude and the real phase:
\begin{equation}
\label{A-phi}
\left\lbrace
\begin{array}{l}
\dfrac{d\rho}{dt}=\dfrac{\rho}{2} \left(\beta-a -\dfrac{b\rho^{2}}{4\omega_{0}^{2}}-bz_{0}^{2}\right),\\
\\
\dfrac{d\phi}{dt}=-\dfrac{b\beta}{2\omega_{0}}\left( \dfrac{\rho^{2}}{4\omega_{0}^{2}} +z_{0}^{2}\right).
\end{array}
\right.
\end{equation}

In the considered range of the parameter $\beta$ values the oscillation frequency $\omega_{0}$ is close to unity: $\omega_{0}=\sqrt{1-\beta a}\approx 1$ (it comes from the first equation of the system (\ref{osc})). The problem of the periodic motion existence and bifurcations in the system (\ref{osc}) is reduced to amplitude equation analysis. The amplitude equation of the system (\ref{A-phi}) has two equilibrium solutions. The first solution, $\rho_{1}= 0$, corresponds to the equilibria $x=0$, $y = 0$, $z_{0} \in (-\infty;\infty)$. When the parameter $\beta$ passes through the value $\beta=a$, the solution $\rho_{1}$ is stable for $|z_{0}|>\sqrt{\frac{\beta-a}{b}}$ and unstable for $|z_{0}| < \sqrt{\frac{\beta-a}{b}}$. The second solution: 
\begin{equation}
\rho_{2}=2\sqrt{\dfrac{\beta-a}{b}-z_{0}^{2}}
\label{rho2} 
\end{equation}
appears at $\beta \ge a+z_{0}^{2}b$, is stable and corresponds to non-isolated closed curve appearance in the vicinity of the equilibrium points with the coordinates $\left( x=y=0, |z|=\sqrt{\dfrac{\beta-a}{b}}\right)$. It should be noted, that the frequency of oscillations corresponding to motions along non-isolated closed curves depends solely on the parameter values. This is due to the fact that the formula for $\dfrac{d\phi}{dt}$ (the second equation of the system (\ref{A-phi})) corresponding to the solution $\rho_{2}$ (Eq. (\ref{rho2})) becomes: $\dfrac{d\phi}{dt}=-\dfrac{\beta}{2}(\beta-a)$.

\begin{figure}
\begin{center}
\includegraphics[width=1.0\columnwidth]{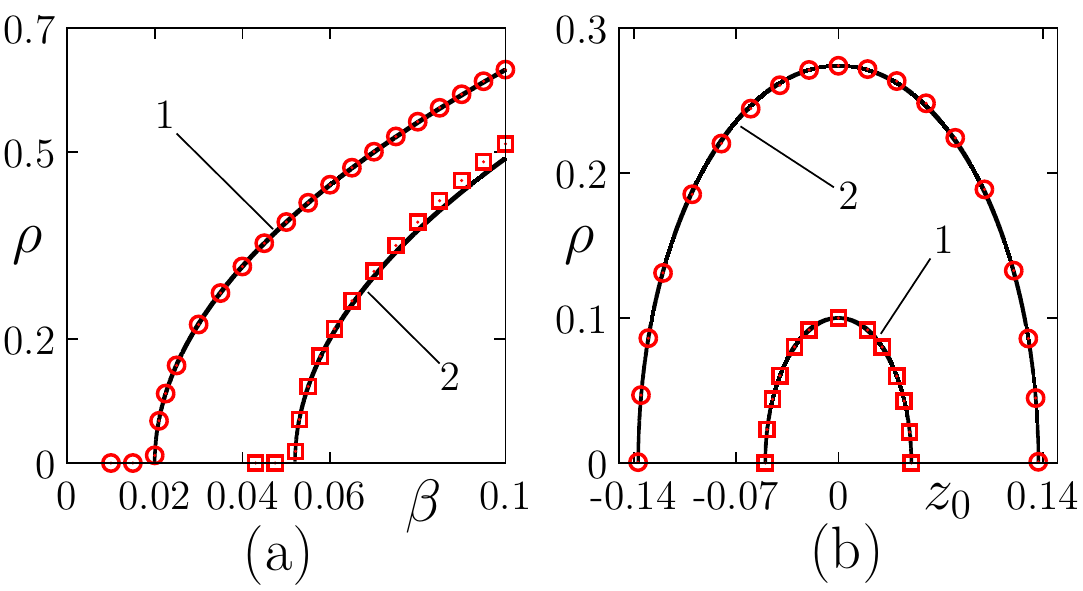}
\end{center}
\caption{System (\ref{system}). (a) Dependence of amplitude, $\rho$, of oscillations $x(t)$ on parameter $\beta$: analytically derived formula (\ref{rho2}) for $z_{0}=0$ (the black curve 1) and $z_{0}=0.2$ (the black curve 2) and registered in  numerical experiment amplitude of oscillations $x(t)$  corresponding to initial conditions in the neighborhood of the origin ($x_{0}=y_{0}=z_{0}=0$) and the point ($x_{0}=y_{0}=z_{0}=0.2$) (the red circles and squares correspondingly). (b) Dependence of amplitude, $\rho$, of oscillations $x(t)$ on initial condition $z_{0}$: analytically derived formula (\ref{rho2}) for $\beta=0.022$ (the black curve 1) and $\beta=0.035$ (the black curve 2) and registered in  numerical experiment amplitude of oscillations $x(t)$  corresponding to initial conditions ($0,0, z_{0}$)  (the red circles and squares). Other parameters are: $a=0.02$, $b=0.8$.}
\label{figure3}                                                                                                   
\end{figure}

Consequently, the following picture of bifurcational changes is revealed. Increasing of the parameter $\beta$ gives rise to loss of stability of the equilibria with coordinates ($x=y=0, z_0$), for which the condition $|z_{0}|<\sqrt{\dfrac{\beta-a}{b}}$ is satisfied. When the point of equilibria losses the stability, a non-isolated closed curve appears in its vicinity. The amplitude of the oscillations $\rho=\rho (\beta, z_{0})$ corresponding to motions along the non-isolated closed curve depends both on the parameter $\beta$ and the coordinate $z=z_{0}$ (Eq. (\ref{rho2})) and increases gradually starting from zero with the parameter $\beta$ growth (Fig. \ref{figure3}(a)). Such scenario of soft oscillation excitation in the system (\ref{system}) is similar to the supercritical Andronov-Hopf bifurcation corresponding to self-sustained oscillation excitation in nonlinear dissipative systems with a finite number of isolated equilibrium points. Moreover, the dependence of the periodic solution amplitude as a square root of supercriticality (as in the system (\ref{system}), see Eq. (\ref{rho2})) is typical for systems, realizing the supercritical Andronov-Hopf bifurcation. Concurrently, the derived dependence $\rho(\beta, z_{0})$ (Eq.(\ref{rho2})) shows that one can realize the bifurcation in a case of a fixed value $\beta$ by changing the coordinate $z$. (Fig. \ref{figure3}(b)). Therefore the bifurcation observed in the system (\ref{system}) is also a bifurcation without parameter. There is the second distinction from the classical Andronov-Hopf bifurcation. The periodic oscillations in systems with a line of equilibria are not true self-sustained oscillations in full sense, because the non-isolated closed curves are not attractors in themselves \cite{korneev2017}. 

\section{Conclusions}
\label{conclusions}
Bifurcational mechanism of the periodic solution appearance in a system based on the memristor with cubic nonlinearity has been analytically revealed. Correctness of  theoretical results has been confirmed in numerical modeling (Fig. \ref{figure3}). Soft excitation of the oscillations consists in the non-isolated closed curve emergence from become unstable equilibrium points belonging to the line of equilibria. Oscillation excitation can be achieved by changing of a parameter as well as by tuning of initial conditions. Peculiarities of the soft oscillation excitation is determined by memristor nonlinearity and has a similar character as compared to the supercritical Andronov-Hopf bifurcation corresponding to soft excitation of the oscillations in nonlinear dissipative systems with a finite number of isolated fixed points. 

Usually bifurcations caused by changing of the parameters and bifurcations without parameters are considered separately. In the present work these two kinds of bifurcations are united in the same phenomenon.  

\section{Acknowledgments}
This work was supported by DFG in the framework of SFB 910 and by the Russian Ministry of Education and Science (project code 3.8616.2017/8.9).
We are very grateful to professor Tatiana E. Vadivasova for helpful discussions.


\end{document}